\documentclass[10pt]{article}
\usepackage{graphicx}
\usepackage{rotating}
\usepackage{adjustbox}
\begin{document}

\begin{center}
\Large\bf{A novel alternative to analyzing multiple choice questions via
  discrimination index}

\end{center}

\begin{center}
P.K.Joshi$^1$, Y. Jain$^2$,  R.Khunyakari$^3$ and S.Basu$^4$

$^1$HBCSE-TIFR, Mankhurd, Mumbai 400088, India.\\
$^2$Department of Computer Science, IIT-Bombay, Powai, Mumbai  \\
$^3$TISS-Hyderabad, Hyderabad, India \\
$^4$ IIT-Guwahati, Guwahati, India

\end{center}

\vspace{0.3in}

\begin{minipage}{4in}

\begin{center}Abstract\end{center}

The value of multiple choice questions (MCQs) in seeking large-scale, high-stakes, educational assessment is widely established. Students' responses to test items with a multiple-choice question format enable assess the extent of students' understanding and also help make valuable decisions about the quality of questions that make robust assessments possible. The use of discrimination index (DI) to analyse MCQs is also widely prevalent in literature \cite{kelly}. This paper makes a case for using a novel approach to analyzing data using the DI. The case for novelty is argued through an empirical, comparative analysis on three sets of data: conjecture data, data from an exam for screening talented students for a competitive event (two examples), and data from an international competitive academic event. The scheme is developed to handle the data gathered from different question formats such as MCQs, Long answer questions (LAQs) and a combination of these two question formats. A code has been developed for carrying out computational analysis on large data sets. A comparison with the conventional approach to data analysis establishes the worthiness of ideas proposed for making meaningful inferences and simultaneously renders it possible to attend to nuances that are greatly compromised while analyzing huge data-sets. The paper brings a critical value-addition to the body of analytical knowledge building.
\end{minipage}

\section{Introduction}
\label{int}

Measurement and assessment have been widely acknowledged as important tool to gaining a certain objective understanding about learning \cite{bloom,haladyna}. While acknowledging that tests and assessments come in many different forms,  \cite{Linn} asserts the value of these in establishing accountability systems intended to improve education. Historically, tests and assessments, he argues, have been key elements in educational reforms. 
Tests, as an assessment tool, have been an integral part of the process of screening the academically talented students from 
socio-culturally rich and linguistically diverse population, who come from different academic exposures and nurture environments.  
\cite{bloom} draws our attention to the significant role of testing in education but at the same time cautions us to deal with it, keeping in mind the purpose of assessment.

\subsection{Centrality of `purpose' in assessments}
\label{purpose}

The `purpose' is integral to the design and selection  of an assessment tool. In the literature on educational assessments, scholars have long argued for the salience of purpose in making crucial educational decisions. Assessments are variously designed to meet what  \cite{tanner}, have broadly categorized as the {\it communicative, managerial} and {\it pedagogical} purposes of assessments. Interestingly, the use of tests reported in this paper draws from a context where test, as a method to assess students' performance, gels with all the prepurposes identified. Newton \cite{newton}draws our attention to prevailing confusion about `assessment purpose' in literature and identifies three distinct interpretations that concern judgement, decision and impact levels. In this paper, the `assessment purpose' being discussed concerns the decision level in Newton's taxonomy, which is about the use of an assessment to support a selection decision for entry to higher education or screening talent. Given this understanding, it becomes really important to develop appropriate assessment tools that effectively match the purpose identified. According to  \cite{cunningham}, meeting four requirements can accomplish good assessment: (i) a knowledge of correct procedures for student assessment; (ii) a commitment of an adequate amount of time; (iii) planning; and (iv) a reflective analysis of the assessment process. 

Careful consideration of all the 4 requirements contribute to validity of test instrument and to a fair degree of realizing the 
intended purpose of assessment. In other words, a sound test instrument is a significant milestone in realizing purpose of assessment. The development of test instruments is an involved process covering all the four requirements outlined by Cunningham. 
A brief on the process of development of the test instrument being used will be discussed in section \ref{dsa}.

Large stake assessments often use items that are largely multiple choice questions (MCQs). They also cover a wide range of students with different cultural backgrounds and orientations. Although MCQs have their limitations, they continue to remain a reliable means to operationalising large-scale, high-stakes, summative educational assessments. Careful and systematic use of these tests help us to gain insight into student's thinking and quality of questions allowing scope for widening the educational value of assessment. In attending to this conceptualization, assessment through test meet the otherwise segregated relation of assessment and learning represented through the classical preposition `of' and `for' learning. Assessment conceptualized {\it as}, {\it of} and {\it for} learning allows not only gauge quality of learning but also enables develop a reflective introspection on test items that could provide opportunities for conceptual processing and developing refined understanding.

\subsection{Item Response Theory in Assessments}
\label{intro}

It is indeed hard to miss the socio-historic strides made in transitioning from the use of Classical Test Theory (CTT) to Item Response Theory (IRT), particularly in situations seeking screening of talent from a pool of respondents, in reliable and meaningful ways from large data sets. The framing of MCQs is challenging as the range of choices need to be designed to capture the likely range of responses, from which the respondent select the most appropriate response. A test which includes such items is amenable to coding, computation and quantitative analysis. However, much investment is needed in developing authentic test items as well as in mining the enormous data appropriately to meet its intended purpose. The efforts towards developing guidelines that aid multiple choice item writing \cite{cohen,haladyna} and evaluating the worthiness of number of choices that afford sufficient enough
distracters to bring out the cognitive challenge in a MCQ  \cite{tarrant} are some areas of contemporary research interest.

While test as a tool to elicit patterns from large data set is irrefutable, the quest for seeking analytical alternatives for drawing precise and subtle insights is an ongoing one. In the context of discussing the best suited approach for analyzing test with single best response MCQs,   \cite{ding} assert that the choice of approach is guided by the purposes and data interpretation. Item analysis not just reveal about patterns in student responses, but also enables assessing the quality of items in the test. Literature   \cite{haladyna,baker}) does demonstrate consensus on how features of the design of test items contribute to the performance on test assessments. For instance, the role of language in designing the stem of the item \cite{paxton}, the appropriate choice of distracters  \cite{vyas}, and the larger purpose or intent of the item itself. Attending to these features add rigor to the insights and inferences one draws from the test assessments. Another feature, that could add value is the understanding of the scheme for analyzing and inferencing from the data-set. It is this feature that has been the central point of discussion in this paper.

As we know, large-scale assessments often use tests as instruments, which largely comprise of single option, multiple-choice questions (MCQs). Typically, each MCQ item includes a stem which details the context and the question and is followed by four response options, of which, one of them is the most appropriate response. In addition to the form and structure of the test item, the time allocated, the marking scheme with relative weightage, guessing behavior, negative (penalty) marking, etc. are known to influence students' performance on the tests. While analysis of students' responses to MCQs reveals response trends among a large pool of respondents, the responses garnered can help make informed judgements about the quality of test items \cite{hassan}, assimilate an understanding of appropriate, plausible distracters that challenge students' thinking \cite{vyas}, and if meticulously developed, can even enrich the conceptual understanding of respondents  \cite{hutchinson,mccoubrie}. The prospects with test items need systematic and careful engagement with data. 

Both in terms of the wide number of students involved and the challenge of screening talent, the MCQs format of test items are reliable and convenient in achieving the larger objective. 
The forms of test items, other than MCQs, are equally important. The other forms include long answer or conceptual questions (numerical or descriptive in nature) and items designed to assess understanding and performing science experiments.  Long answer questions (LAQs), especially of numerical calculation type, are a tool to understand the ability of the student to extract the science principles hidden in question, in order to solve the problem. Tests, in these cases, can be reliably used to gauge the ability of students to handle multiple concepts, ability to read a question carefully, follow an experimental procedure meticulously and several such student attributes. The LAQs can be experimental tasks which inquire into the practical, hands-on capacities of students. Besides discussing the scheme to analyze the MCQs, this paper also opens up the possibility of expanding the scheme to LAQs, which have seldom been explored in literature.

\section{Situating the study: Motivation and context}
\label{mot}

The motivation for this study stems from two important standpoints: a conceptual tension and an empirical observation drawn during analyzing a data-set. Both standpoints contributed to problematising the work being reported in this paper. The process of developing an understanding of logic in the analytical procedure helped us reach an alternative that could reliably address the issues at hand. Let us develop an understanding of the problem in focus in this study through the means of an elaborate example. 

The Item Response Theory (IRT) is a widely used theoretical framework for analyzing MCQs. The IRT identifies Difficulty Index (Diff-I) and Discrimination Index (DI) as two critical constructs which reveal about the quality of test items. This section briefly discusses the algorithms used to arrive at these indices in the conventional analytical approach. 

Let us define an examination where there are N students attempting P questions. Let  student S$_i$, (i=1,N) appear for the examination which has questions Q$_j$ (J=1,P). The students are listed, sorted by the total marks in decreasing order. To calculate the discrimination index (DI) for any j$^{th}$ question Q$_j$ the top 27$\%$ of the students are identified and similarly the bottom 27$\%$ of the students are identified \cite{kelly}. For the purposes of this article, let us name them as the ``top block''. Let N$_t$ is the number of students in the ``top block'' list who have 
answered the question correctly. Let
N$_b$ be the number of students in the ``bottom block'' list who have answered the question Q$_j$  correctly. Those students who fall in the list, between these two blocks represent the ``middle block''. Let us define quantity C = 0.27 x N and M$_j$ 
as the maximum possible marks for the j$^{th}$ question.
  
Then, the discrimination index (DI) is defined as: 

\begin{equation}
DI_j ~=~ \frac{N_t - N_b}{{\rm C }} \label{di-def}
\end{equation}

This has been the conventional way of defining and arriving at DI. Similarly, the difficulty index (Diff-I) is defined as the number of students who have answered a particular question correctly, in percentage.
Researchers define difficulty index (Diff-I) as:

\begin{equation}
\frac{N_t + N_b}{ 2 ~ {\rm C}}
\end{equation}

In carrying out item analysis, the scores on the two parameters of difficulty index (Diff-I) and discrimination index (DI) are used to evaluate the quality of test items \cite{kelly}. Those questions which are found to have a lower or negative discrimination index are considered to be not of good quality. Here, good quality implies that the questions are desirable in terms of concept clarity and presentable with good distracters for the sake of discrimination and the purpose of selection in the exam. Literature recommends that the items which discriminate poorly need to be inspected for possible deficiencies and rectified or weeded out completely \cite {hambleton,baker}. Thus, the extent of discrimination of a test item within a test instrument plays a decisive role in making judgements about the quality of an instrument for assessment. This understanding brings to the conceptual motivation driving this study, which led us to inquire whether there is a pattern in the kind of questions that have received varying numbers on the discrimination index. In other words the design or sequence of questions or selective non-response pattern may lead to penetration of a bias. In conventional analysis, the relation between the two parameters of Diff-I and DI is ascertained through Pearson coefficient(r), which further contributes to developing a sense of the relation between the two constructs \cite{cohen}. However, the analysis reported in this paper does not focus on the use of Pearson coefficient(r).

The empirical motivational stand-point can be elaborated with an illustrative example drawn from our experience with a data-set. As we engage with the data-set, the example illuminates the kind of questions that may arise in the process of analysis and pin-points the areas of unrest in subscribing to the conventional analytical approach.

\subsection{Relating through an illustrative example}
\label{sec:Present-work}

Let us take an example of 290 students with 30 questions with Diff-I and DI values derived through conventional analysis, in Table \ref{tab-injso-2017}. This is the data of sample 1 also discussed in section \ref{data-injso-2017}.

\begin{table}[h]
\begin{center}
\caption{Table of DI for the sample exam described in text}\label{tab-injso-2017}
\begin{tabular}{|c|c|c|c|c|c|c|c|c|c|c|}\hline
Q. No & 1 & 2 & 3 & 4& 5 & 6& 7 & 8& 9 & 10   \\ \hline
Diff-I &.241 & .08 & .75 & .57 & .64 & .27& .39 & .35 & .41 & .30  \\
DI & .19 & -0.03 & .21 & .35 & .38 & .32& .27 & .22 & .49 & .35  \\ \hline
Q. No & 11 & 12 & 13 & 14 & 15 & 16 & 17 & 18 & 19 & 20 \\ \hline
Diff-I& .67 & .62 & .82 & .25 & .35 & .31 & 1.0 & .44 & .13& .73 \\
DI & .45 & .57 & .12 & .12 & .4 & .22 & .00 & .16 & .1& .42 \\ \hline
Q. No   & 21 & 22 & 23 & 24 & 25 & 26 & 27 & 28  & 29 & 30  \\ \hline
Diff-I  & .31& .35 & .73& .33 & .15 & .55 & .83 & .40 & .44 & .49 \\
DI  & .29& .27 & .27 & .23 & .01 & .22 & .18 & .21 & .23 & .19  \\ \hline
\end{tabular}
\end{center}
\end{table}

\begin{figure}[h]
\includegraphics[scale=0.4]{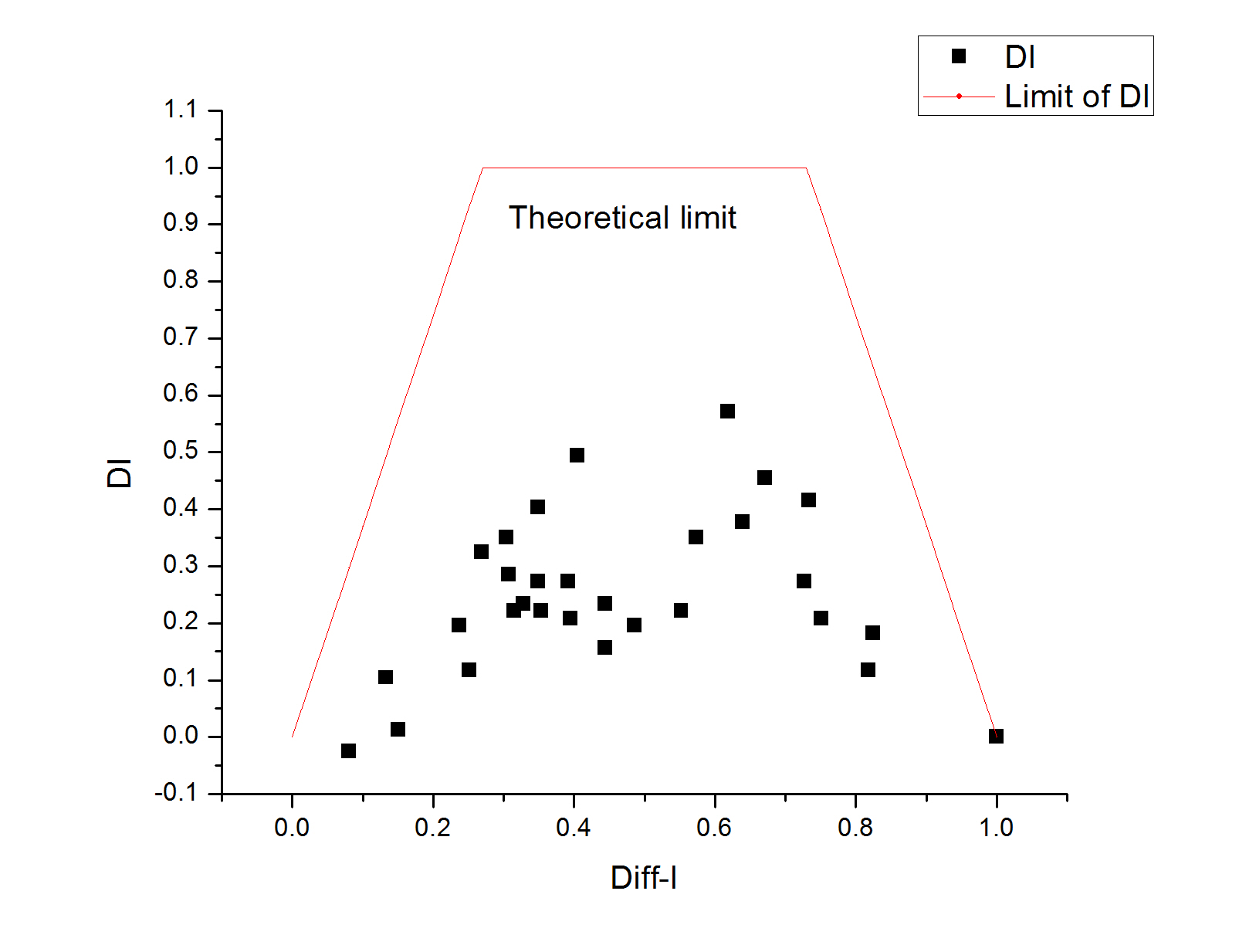}
\caption{Relationship between Diff-I  and DI}\label{di-di}
\centering
\end{figure}

In Figure \ref{di-di}, one can see the relationship between the number of students who have attempted the question correctly with the DI. It can be noted that if a question is answered correctly by all the students, difficulty index (Diff-I) of 1.0, then discrimination index (DI) will have a value of 0.0. On the other side of the spectrum, if no student has answered a question correctly, Diff-I of 0.0, then again DI will be 0.0. If very few students have answered the questions correctly, then the DI values will be very close to 0.0 and can even have a negative number as in the case of question item 2 in our sample. Only when Diff-I is more than 0.27 and less than 0.73, the DI value can have a value of 1.0. For all other values of Diff-I, upper limit of DI will have values between 0.0 and 1.0. The theoretical limit is plotted in the Figure \ref{di-di} indicated by a red line. 

The calculation is further called into question, if we are to consider questions that involve negative scoring. A negative scoring implies a 0.25 mark/score deduction for every question item responded incorrectly. The items attempted by respondents with an element of ``adventure" or ``risk" in attempting maximum number of questions is curtailed by imposing a negative marking scheme. Hence, assuredness in understanding and responding to questions is noted to be more rigorous in use of tests that involve a negative marking scheme. If we were to consider negative marking, in the equation \ref{diffi-equn} the numerator has an additional term (0.25 x total number of students) and the denominator has maximum marks ($M_j$) as 1.25 . This
will be dealt with at approprite relevance in later section. 
The theoretical limit, for the j$^{th}$ question, is calculated on the basis of Diffi-I in three phases. 

For questions whose Diff-I is less than 0.27, the limit is

\begin{equation}
\frac{ {\rm ((Diff\!\!-\!\!I) ~x ~100)} }{{\rm C ~x~ M_j}}
\end{equation}

For the questions whose Diff-I is more than 0.27 but less than 0.73, the theoretical limit is 1.0.
 
For questions whose Diff-I is greater than 0.73 it is 

\begin{equation}
\frac{ {\rm ((1~- Diff\!\!-\!\!I) ~x ~100)}}{{\rm C ~x~ M_j}}
\end{equation}

In the conventional method of calculating DI, the ``total'' is defined by adding all the marks for each student, including the marks of the question whose DI is being calculated and occasionally even the marks from questions, which are not in MCQ format may be included. In the example discussed above the negative marking of MCQs was included while deciding the total, for sorting out the list of students and deciding which of the students get included in the ``top block'' and which of the students fall in the ``bottom list'' of the students.

It is also observed that which students fall in ``top block'' and ``bottom block'' are not uniquely defined. Let us explain this by a simple example. Consider 300 students, which implies that 81 students fall in both the ``top block'' and the ``bottom block''. After arranging the students data in the decreasing order of marks, if students who come at position 79 to 84 all have same total marks, then depending on how the list is arranged, only 3 (out of 6) students come in the ``top block'' who have secured the same total marks. This means a different format of sorting the list of students can put entirely 3 different students in the ``top block', thus giving a different value of DI for the j$^{th}$ question under consideration. Depending on the number of students in such category, DI values have been observed to vary significantly. Such observations from empirical data-set attracted our attention to a string of questions that may have serious implications on assessment practices. For instance, is it possible to drop a few questions from the list of `P' questions to calculate the ``total score''? How will it impact the value of DI? How will the value of DI vary, if different combinations of questions are considered to define the ``total'' to calculate DI? More importantly, what is the impact if total is decided on the basis of marks of remaining (P-1) questions? In this paper, this is called as ``without the self marks'', and the conventional method is defined as ``with the self marks'', included to calculate the total.

It has been observed that for several examinations, where negative marking is included, the incorrect responses are not included in the analysis. Thus, the students who left the question unanswered and those who answered it wrongly, get clubbed together into one single category. However, this paper argues the value of making this distinction and the impact of this decision in evaluating the value of DI.

The conceptual and the empirical motivations converged when we noted a certain pattern in scores obtained on a test on a data-set. We found that some questions had a very low DI value and even had a negative sign occasionally had also a corresponding low Diff-I value. This alerted us to look at the nature of questions which revealed a trend. The questions with a negative value were those that probed conceptual understanding. In other words, the item was designed with a deliberate intent to bring out the conceptual understanding or grasp of the respondent. Thus, in a conventional analysis a test item may have been counted as not so good in quality and therefore, would have been eliminated from the test  \cite{hassan,pyrczak}. This made us revisit the technique of item analysis itself.

\subsection{Ascertaining authenticity of motivation using data simulation}
\label{sec:sim}

\begin{figure}[h]
\includegraphics[scale=0.4]{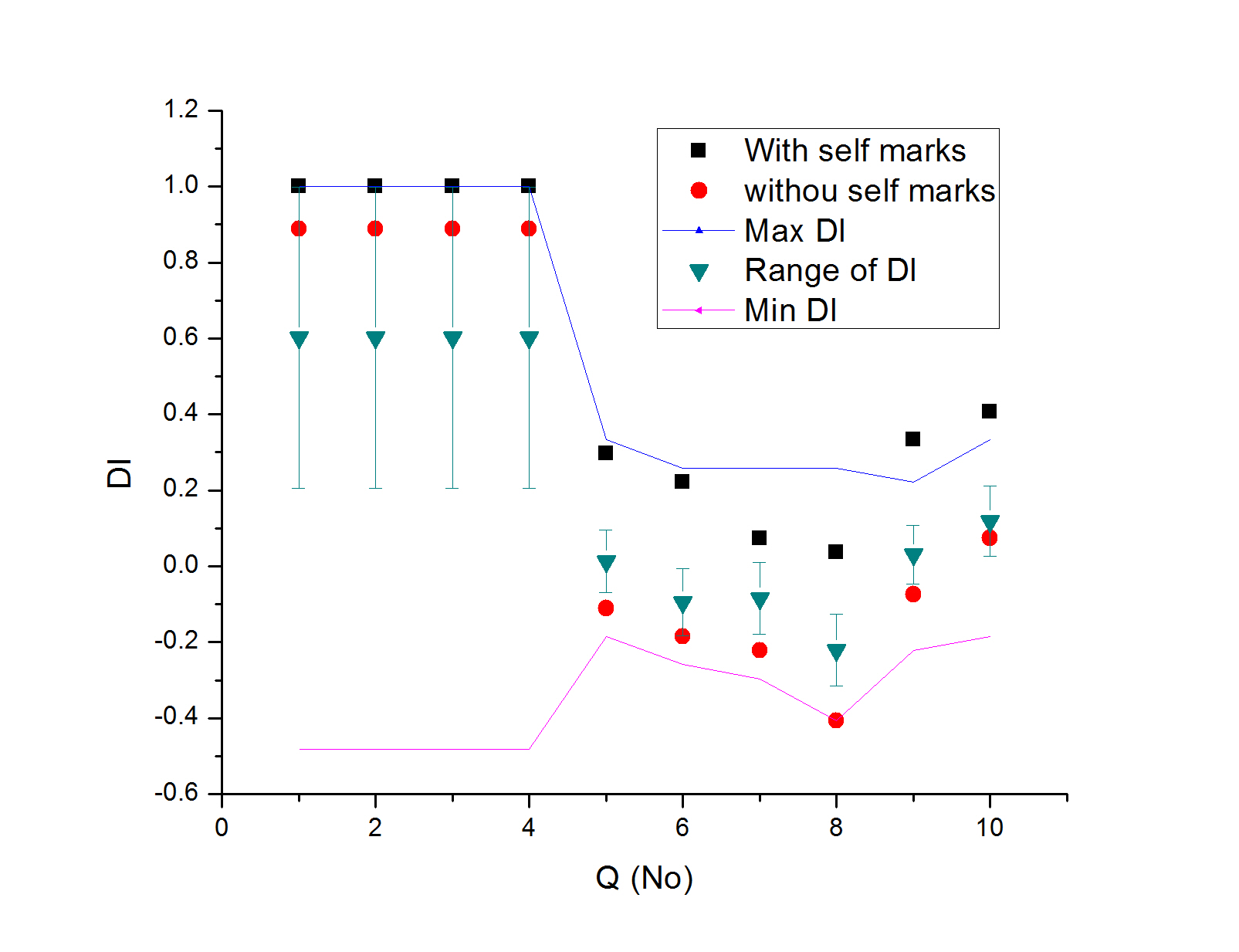}
\caption{Different values of DI as a function of question  number. Here Diff-I is 0.5 for all questions. }\label{manud}
\centering
\end{figure}

A set of data was generated for the purpose of simulation. The set consisted of 100 students and 10 questions. The set was so designed that first 50 students (roll number wise) answered first 4 questions correctly. Whereas for the remaining 6 questions, 50 students answered them correctly but these were some 50 students randomly selected from the set of 100. Here value of C is 27 and Diff-I for all questions is 0.5.
The values of DI calculated can be seen in Figure \ref{manud}

When DI values were calculated the conventional way, it turned out that the DI value for first 4 questions gave 
a value of 1.0 and a high value for the 10$^{th}$ question. When the subset of these 5 questions was separated out, and analysis carried out for all possible combinations of these 5 questions, the DI value for question number 10 dropped to a low value whereas the DI for the first 4 questions remained 1.0. These values can be seen in Table 
\ref{manud-tab}.

\begin{table}[h]
\centering
\caption{Table of average DI with standard deviations for questions 1,2,3,4,and 10}\label{manud-tab}
\begin{tabular}{|c|c|}\hline
Q. No  &  Average  DI                           \\ \hline
1 &  0.9 $\pm$ .3  \\
2 &  0.9 $\pm$ .3  \\
3 &  0.9 $\pm$ .3  \\
4 &  0.9 $\pm$ .3  \\
10 &  0.11 $\pm$  ($<$) .01  \\ \hline
\end{tabular}
\end{table}

Thus, isolating the subset of first 4 questions separate from the remaining 6 questions. This data was generated to test the newly developed computational method and test if ``grouping'' is possible in-principle. The lurking question, then becomes, 
is it possible to identify a subset of questions which all have very high values of DI? 

\begin{figure}[h]
\includegraphics[scale=0.40]{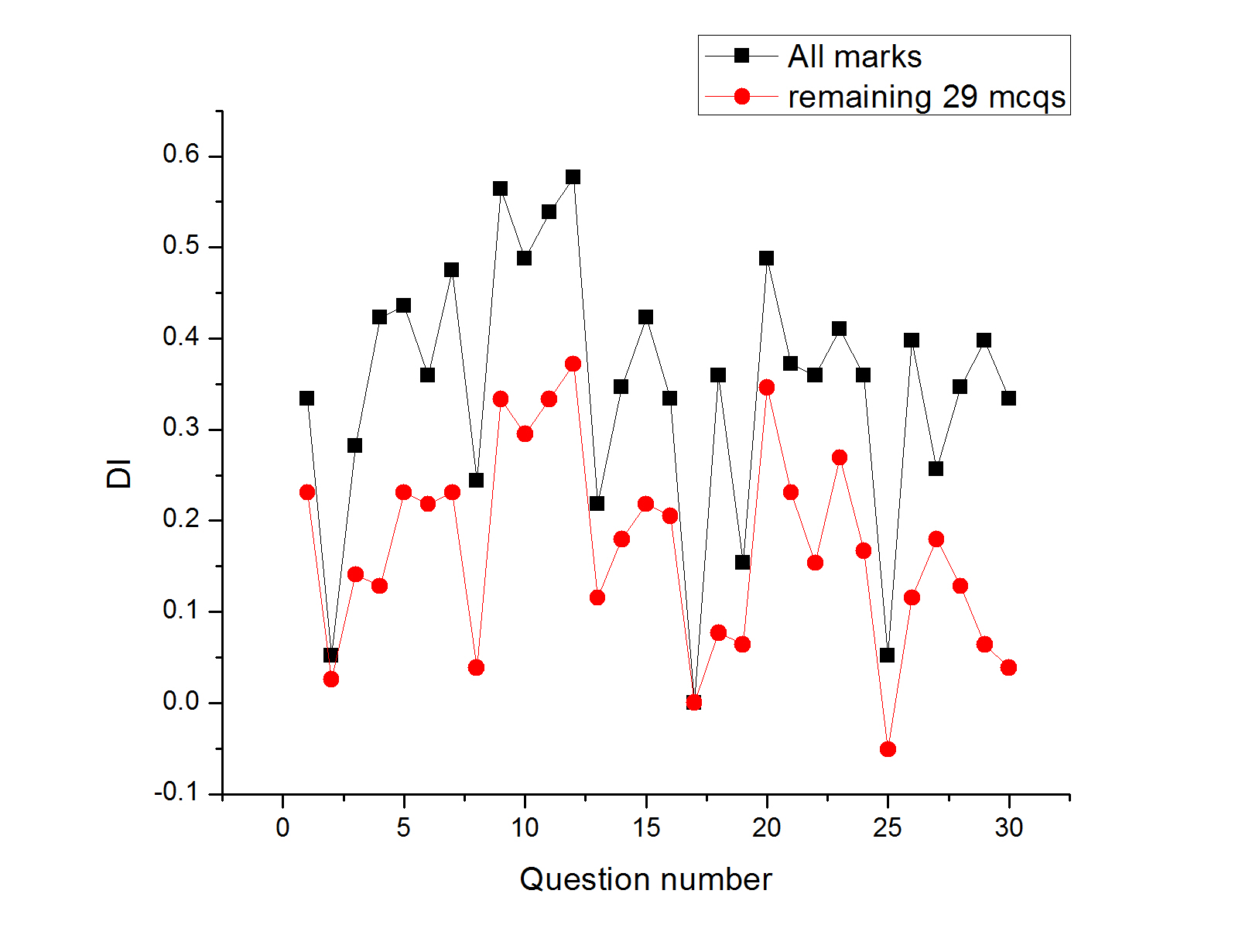}
\caption{Difference between the DIs on the basis of total when the marks of all questions (including negative marking)
 are included and when the marks of remaining 29 questions are considered to calculate the total. This is from first example discussed in the paper, i.e. INJSO 2017.}\label{with_without}
\end{figure}

\section{Objectives of the study}
\label{obj}

This study aims to revisit the prevalent, conventional approach to analyzing multiple-choice questions using the constructs of Diff-I and DI. Using an empirical, data-driven approach, the study seeks to compare the findings obtained from the computed data-sets and bring to discussion some gray areas, which perhaps can be strengthened with suitable alterations. 
The objective has been systematically addressed through raising research questions that bring greater focus and direction to the work. 

The research questions that guided this study are as follows:

\begin{enumerate}
\item What kind of questions remain unanswered while applying the conventional approach (prevalent algorithm) using the item response theory framework? 
 
\item How can the existing algorithm be orchestrated to meet the demands of rigor and insight from the analysis of diverse question formats (such as, multiple-choice questions, long-answer questions) employing negative (penalty) scoring scheme? 

\item What kind of challenges and prospects does the alternative algorithm present?

\item Do the experiences of applying the alternative algorithm establish its worth with large data-sets and different test item formats?     

\end{enumerate}

\section{Description of the study}
\label{des}

The study is a data-driven, empirical effort to introspect into conventional procedures for analyzing and interpreting response trends on test items using the IRT. The study makes an effort to suggest a novel alternative to strengthening inferences gained from analyzing multiple choice questions. The proposed alternative is derived and tested on three large data sets, involving student scores on test items. The computational analysis arrived at using the alternative procedure provides a neat and elegant way to use the notion of discrimination and difficulty indices not just to adjudge the quality of performances but also as a means to reflect on quality of questions. Besides this, the alternative procedure promises to develop strong grounds for noticing and minimizing several effects that come into play in interpreting results. These include assessing meaningfully the impact of no-responses to certain items, guess behavior patterns in responses or negative (penalty) marking being employed. Also, the study goes beyond the conventional limiting of analytical procedure to MCQs for test analysis.The study extends the same analytical scheme to LAQs.

\subsection{Study methodology} 
\label{meth}

The study methodology is located within the quantitative research paradigm. The thrust of the work reported, in our opinion, falls within the analytical research tradition. The study reported begins from the large-scale data being analyzed to assess students' performance on a test that is used to screen and select talented students. Our engagement with literature on the process of analyzing data led to identifying and problematising a conceptual area of discomfort which, in principle, could lead to inferences that need to be treaded with caution.

Through the process of analysis and discussion, a novel scheme has been envisaged which has been experimented with three data sets, reported in this paper. DI values are calculated using different techniques deviating from the conventional techniques. This exercise was carried out to study some of these variations and observe the changes in the value of DI.

\subsection{Sample and the data}
\label{dsa}

The data constitutes the most significant {\it subject} and {\it object} examined in this study. Primary data being used has been gathered through authentic sources of large-scale, assessments. This section briefly describes the nature of data-sets used along with the rationale for its use in this study. 

Three sources of primary data have been consulted in this study. All data pertain to test scores of students in the age group 14-16 years, pursuing or participating in an established International, academic competition in the sciences at the secondary grade level. The academic, competitive event is an annual feature that invites student contingents, participating from across the globe to participate in what is referred to as the International Junior Science Olympiad (IJSO) (for further details, see www.ijsoweb.org). In IJSO competition, the questions designed are challenging in nature and are approved by the International Board of IJSO which consists of teachers, educationists from all the participating countries. Test items, which are a part of the test papers, are designed to be intellectually challenging and to serve the purpose of screening talent from student resource pool. In terms of the intent, the tests thus serve the dual purpose of {\em diagnostics} and {\em summative testing}, identified by \cite{cohen} . Given the purposes, these tests include items which assess knowledge comprehension, reasoning abilities, conceptual clarity, and application of knowledge. On many occasions more than one of these purposes are targeted through a test item. 

The three data sets used in the study represent students' performance scores on tests from: (a) preparation training in India for an international, academic competitive event in 2017; (b) preparation training in India for the international event in 2016; and (c) the international event organized and hosted in India in 2013. While (a) and (b) are trainings organized in India towards preparations for international event, these are referred to as Indian National Junior Science Olympiad (INJSO). The outcomes of the tests result in selection of a contingent representing India for the international event. The international event or IJSO was organized and hosted by India in 2013, hence data set represented as (c) comprises of IJSO 2013 data. It is important to draw attention to the fact that the three data sets involved three different sample of students and also different sets of test items. The process of setting the test items and test validation from the experts for all the examples above are consistent. A comparison of algorithm on data involving three distinctive sets helped ascertain the relevance and worth of the proposed algorithm. More importantly, it allowed scope for experimenting worthiness of the proposed algorithm with other question formats. The algorithm proposed helped handle data gathered from different question formats such as MCQs, LAQs and a combination of these two question formats.

Since large data sets were involved, a computational program was designed for running large data sets by one of the authors of this paper. The outcomes were cross-verified through random checks. The analysis of data sets revealed interesting insights into measurement and assessment, which will be discussed elaborately in the next section.

\section{Working with data: Analyzing and Interpreting }
\label{aid}

This study examines data sets and raises important questions about educational measurements that have implications for assessment of students' performances. One of the first endeavors in this study developed from an understanding of applying the established, conventional approach to analysis of data set. This led us to raise questions that seem to remain unanswered within the purview of existing algorithmic, statistical procedure in use  for analysis. Some of these questions raised were: Are the performance outcomes and values of DI and Diff-I reliable in a case where students have differential patterns of responses to test items or in cases where students have not attempted a few items? Would the DI and Diff-I values vary for tests with different numbers and kinds of question formats?  

Analysis revealed to us that the present methodology of calculating the DI values have certain lapses which have been explained below.  

i. Assume a case with 300 students and 30 questions, resulting in value of C (0.27 x total number of students) = 81. When the data is sorted as per the total marks obtained by 300 students, let us assume that students listed from number 75 to 90 have the same total. In which case for all the $N_t$ students the last 8 students can be included in various different ways out of the total 16 students who have received the same total marks. Thus resulting in a varying number of DI values. The proposed alternative in this paper is designed to handle all such discrepancies, since it looks at various different ``total'' values for selecting $N_t$ students. 

ii. In case of analysis of MCQs, if negative marking is included in the assessment of students, then the current methodology clubs the wrong answer and the no-answer together treating them {\it at par}. In literature, it is often argued that negative (penalty) marking helps discriminate between students who answer incorrectly from those who decide to leave it unanswered, in a multiple choice format. We argue and empirically demonstrate that segregating the two categories gives deeper insights into the quality of the questions.

Drawing on the outcomes of data simulation and an exposure to the limitations of the existing algorithm, the question about a possibility of correction or revision of the existing algorithm occurred. Thus, the question now became how can the algorithm be orchestrated to meet demands of rigor and analyze diverse question formats (such as, multiple-choice questions with and without negative markings, long-answer questions etc. ). To include negative (penalty) marking and tasks which are not multiple choice in nature but long questions which may have maximum marks larger than 1.0, the equation for calculating DI has been modified.

Let Max marks for the j$^{th}$ question be  M$_j$. Let the sum total of marks received by the N$_t$ students be S$_t$ and the sum total of marks received by the N$_b$ students be S$_b$ then new definition of the discrimination index (DI) is 

\begin{equation}
DI_j ~=~ \frac{S_t - S_b}{0.27  ~ {\rm x } ~  N {\rm x} ~ M_j} ~=~ \frac{S_t - S_b}{ C {\rm x} ~ M_j} \label{di-def-new}
\end{equation}

where, it can be clearly seen that in the multiple choice type questions have M$_j$ =1.0  and hence placing the value of M$_j$ as 1.0 in equation \ref{di-def-new} leads to equation \ref{di-def}. The beauty of the equation is that it can handle negative marking in MCQs without any modification to equation or the data. Thus, an additional feature is introduced in this analysis of MCQ questions where ``wrong answer" and ``blank answer" are not treated {\it at par}. In case of negative marks being included the DI values can vary between 1.25 to -1.25, instead of 1.0 to -1.0 and hence need to be scaled dividing the DI values by 1.25. 
 
Similarly, the difficulty index (Diff-I) is {\bf{ redefined}} as the number of students who have answered a particular question correctly, in percentage. Thus, 

\begin{equation}\label{diffi-equn}
{\rm Diff\!\!-\!\!I }~=~\frac{ {\rm Total~ of~ marks~ received~ by~ all~ students}}{{\rm M_j ~x~ N}}
\end{equation}

which is to be modified for MCQ  with negative markings as

\begin{equation}\label{diffi-equn-nm}
{\rm Diff\!\!-\!\!I }~=~\frac{ {\rm Total~ of~ marks~ received~ by~ all~ students} + (0.25 {\rm x ~N})}{{\rm M_j ~x~ N}}
\end{equation}

The present definition of Difficulty Index does not exclude any students which overcoming the problem of exclusion in traditional definition.  The next important question that concerned this study was to identify the kind of challenges and prospects that the proposed algorithm presented. Hence, it became important to learn if the experiences of applying the alternative algorithm established its worth with large data-sets. 

\subsection{Data set 1: INJSO 2017}\label{data-injso-2017}

In this section, we try to determine the impact of different definitions of ``total'' on the value of DI of the, j$^{th}$ question, under consideration.  The sum of marks on all questions except the marks of j$^{th}$ question. The result of dropping the marks of the j$^{th}$ question in deciding the total can be seen in the Figure \ref{with_without}. The {\it red} marks are the DI values when the marks of the question, which is being analyzed, are not included in calculating the total. The black points are when negative (-ve) marking  is included and even the marks of the question under consideration are included in deciding the total.  Figure \ref{with_without} indicates large differences in the values of DI.
~~
\begin{figure}[h]
\includegraphics[scale=0.40]{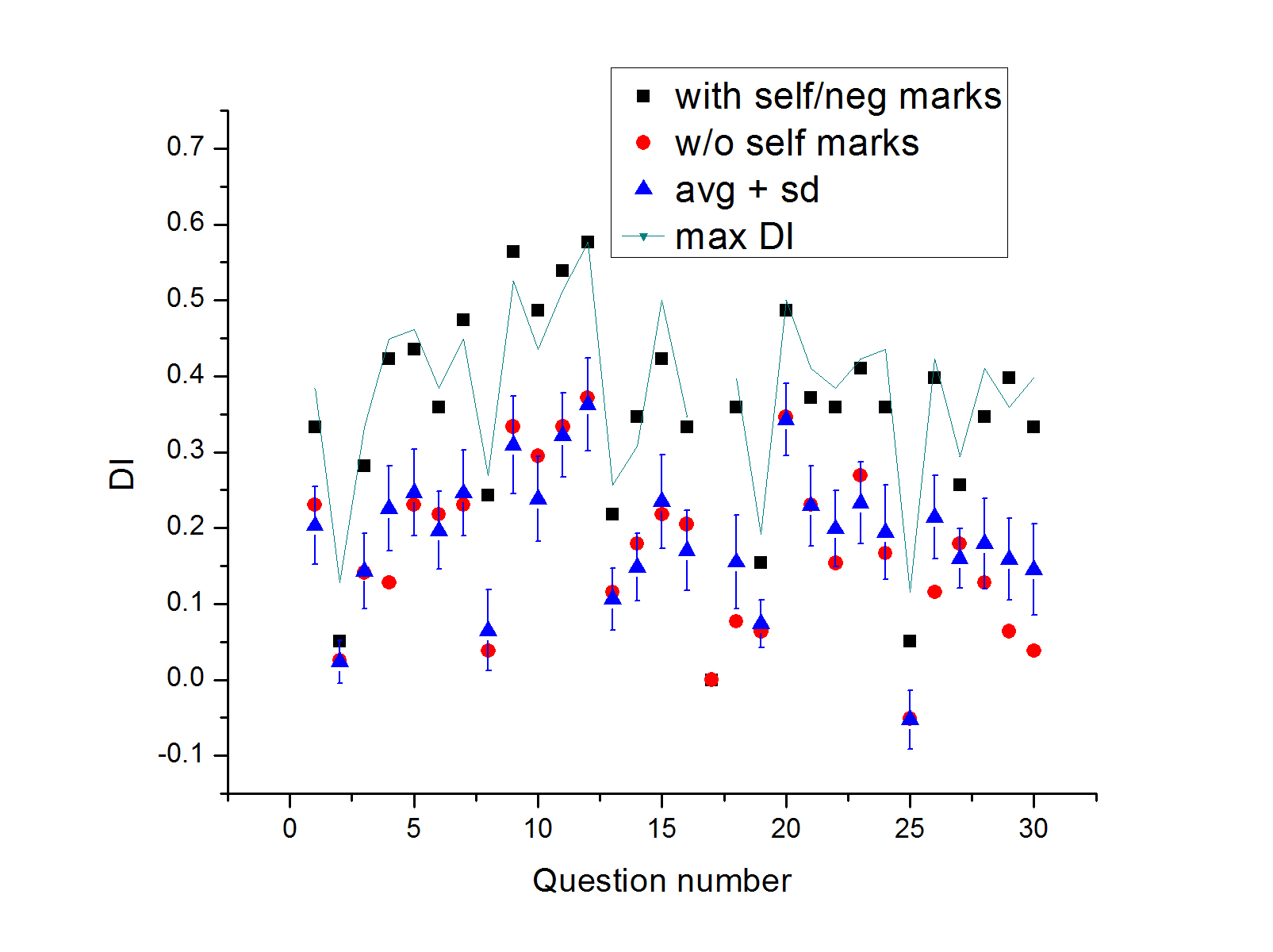}
\caption{Difference between the DIs on the basis of total when the marks of same question is included and not included. This plot represents data from sample1, discussed in section \ref{data-injso-2017} Line representing the maximum DI values is just a visual guide to the reader to compare with other values.}\label{DI_range}
\end{figure}

The ``total" can be further defined as sum of marks of questions which can be different combinations of the set of 29 questions. It was observed that for considering several combinations of more than 8 questions, the number of permutations were so large that it was difficult for common computing devices to calculate them.  If one were
to take all the possible combinations of 2,3,4... 29 questions, the total combinations would be 536,870,882. i.e. 5.37 x 10$^8$ combinations. For 30 questions this becomes 1.61 x 10$^{10}$ trials, which is difficult for a normal laptop computer to handle, hence a random number operation is applied to truncate this large size to a small but manageable smaller sample size which represents this full dataset. The DI values for each question were calculated using around fifty thousand different combinations of the remaining 29 questions. The DI values thus obtained, followed a Gaussian distribution leading to an average value with a standard deviation (covered in more details in the later part of this section). The values of the DI thus obtained can be seen in Figure \ref{DI_range}. The repeatability of the values of DI calculated via this procedure is discussed later in Section \ref{sec-ijso2013} .  From Figure \ref{DI_range}, it can be seen that the DI values obtained when the self marks of the question are not included in deciding ``total'', they fall in the range determined by the average and the respective deviation. However, the DI values obtained when self marks are included in calculating the ``total", and inclusion of negative marking, are much higher and outside the range.

\begin{table}[h]
\centering
\caption{Table of average DI with standard deviations for INJSO 2017}\label{tab-di-range}
\begin{tabular}{|c|c|c|c|c|}\hline
Q. No  &   DI           &         DI       &          Average DI            & DI             \\
           &   With self &  Without self &      With $\sigma$         & Maximum\\  \hline
1	&	0.33	&	0.23	&	0.20	$\pm$	0.05	&	0.40	\\
2	&	0.05	&	0.03	&	0.02	$\pm$	0.03	&	0.13	\\
3	&	0.28	&	0.14	&	0.14	$\pm$	0.05	&	0.33	\\
4	&	0.42	&	0.13	&	0.23	$\pm$	0.06	&	0.44	\\
5	&	0.44	&	0.23	&	0.25	$\pm$	0.06	&	0.46	\\
6	&	0.36	&	0.22	&	0.20	$\pm$	0.05	&	0.37	\\
7	&	0.47	&	0.23	&	0.25	$\pm$	0.06	&	0.46	\\
8	&	0.24	&	0.04	&	0.06	$\pm$	0.05	&	0.28	\\
9	&	0.56	&	0.33	&	0.31	$\pm$	0.06	&	0.53	\\
10	&	0.49	&	0.29	&	0.24	$\pm$	0.06	&	0.44	\\
11	&	0.54	&	0.33	&	0.32	$\pm$	0.06	&	0.51	\\
12	&	0.58	&	0.37	&	0.36	$\pm$	0.06	&	0.59	\\
13	&	0.22	&	0.12	&	0.11	$\pm$	0.04	&	0.27	\\
14	&	0.35	&	0.18	&	0.15	$\pm$	0.04	&	0.31	\\
15	&	0.42	&	0.22	&	0.24	$\pm$	0.06	&	0.45	\\
16	&	0.33	&	0.21	&	0.17	$\pm$	0.05	&	0.35	\\
17	&	0	&	0	&	 		 	&	 	\\
18	&	0.36	&	0.08	&	0.15	$\pm$	0.06	&	0.38	\\
19	&	0.15	&	0.06	&	0.07	$\pm$	0.03	&	0.19	\\
20	&	0.49	&	0.35	&	0.34	$\pm$	0.05	&	0.5	\\
21	&	0.37	&	0.23	&	0.23	$\pm$	0.05	&	0.41	\\
22	&	0.36	&	0.15	&	0.20	$\pm$	0.05	&	0.42	\\
23	&	0.41	&	0.27	&	0.23	$\pm$	0.05	&	0.42	\\
24	&	0.36	&	0.17	&	0.19	$\pm$	0.06	&	0.44	\\
25	&	0.05	&	-0.05	&	-0.05	$\pm$	0.04	&	0.14	\\
26	&	0.4	&	0.12	&	0.21	$\pm$	0.05	&	0.44	\\
27	&	0.26	&	0.18	&	0.16	$\pm$	0.04	&	0.31	\\
28	&	0.35	&	0.13	&	0.18	$\pm$	0.06	&	0.38	\\
29	&	0.4	&	0.06	&	0.16	$\pm$	0.05	&	0.40	\\
30	&	0.33	&	0.04	&	0.15	$\pm$	0.06	&	0.42	\\   \hline
\end{tabular}

\vspace{0.1in}Note: Q17 was dropped due to certain unavoidable reasons

 and hence listed as 0.
\end{table}

The maximum value of DI for any question, obtained from any of the several combinations is sometimes higher than the value obtained when self marks are included to calculate the ``total''.

The average value of the DI for each of the questions is listed in Table \ref{tab-di-range} where it can be seen that the DI values (where self marks are excluded in calculation of ``total") is within the range of values decided by average with uncertainty  or to say within ``the experimental error". The exceptions are question number 4, 18, 26, 29
and 30 which fall in the range of 2$\sigma$.

From the Table \ref{tab-di-range} it can be seen that the maximum DI value resulting from different combinations can even be higher than the value which comes from total based on all the questions, including self. Such questions can be grouped together as a sub-group of questions.  Question 12 has the highest of any DI value which is 0.58 which comes from a combination of group of questions.  In the present work, Question 12 has the highest DI from a group of Question 1, 2, 3, 4, 5, 8, 9 10, 13, 15, 16, 19, 20, 27 and 29. The same can be said in a different way, that if the whole exam had only these fifteen questions, all DI values will be very high value. Now whether such a situation is desirable is a point of further research.

\subsection{Data set 2: INJSO 2016}\label{sec:injso2016}
In the exam conducted in January 2016,
\begin{figure}[h]
\begin{center}
\includegraphics[scale=0.25]{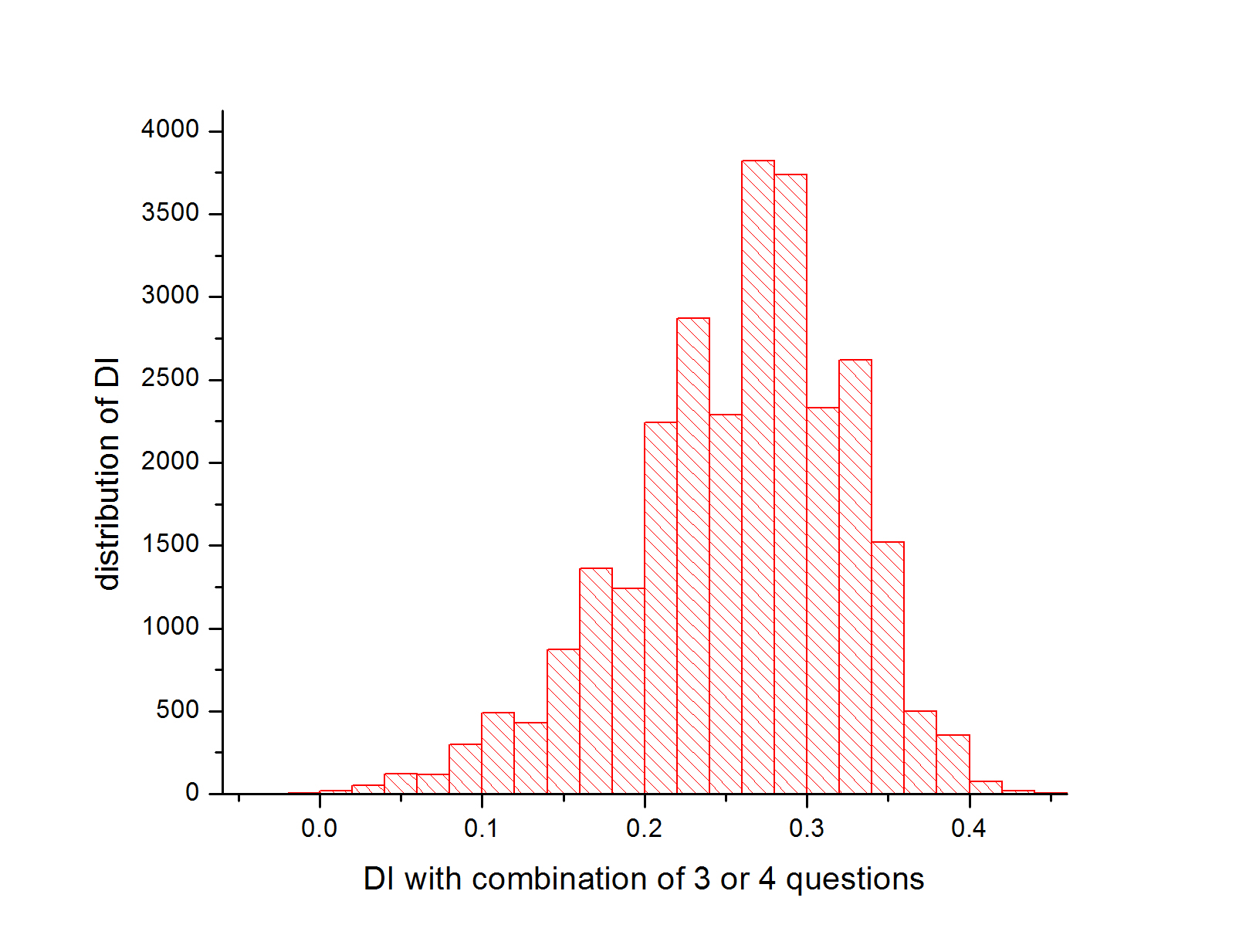}
\caption{Distribution of DI values when only combination of remaining 3 and 4 questions are considered for calculating
total.}\label{sk-gauss}
\end{center}
\end{figure} 
\begin{figure}[h]
\begin{tabular}{@{}l@{}   @{}l@{}}
\includegraphics[scale=0.25, trim=5mm 0mm 40mm 0mm, clip]{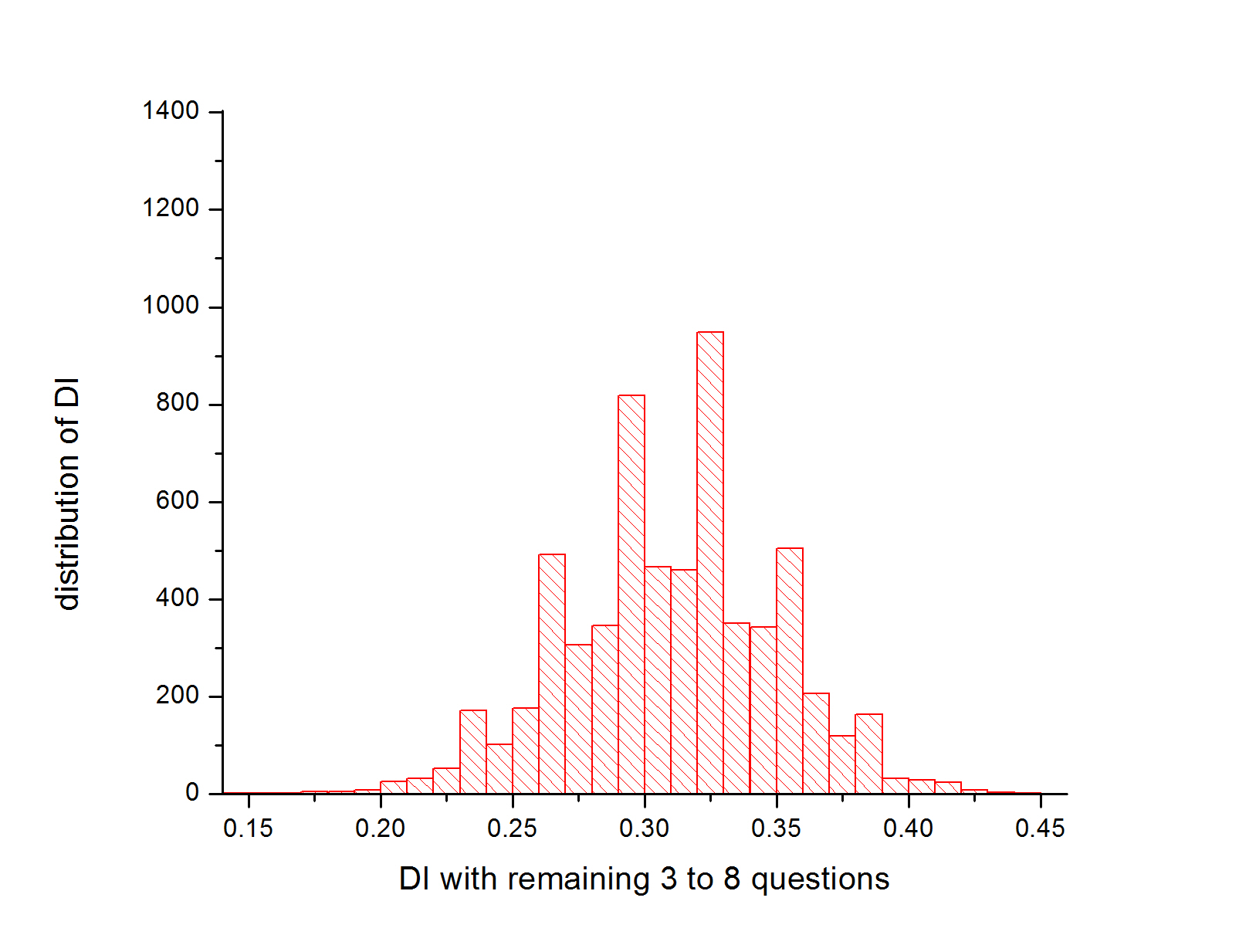} 
\includegraphics[scale=.25]{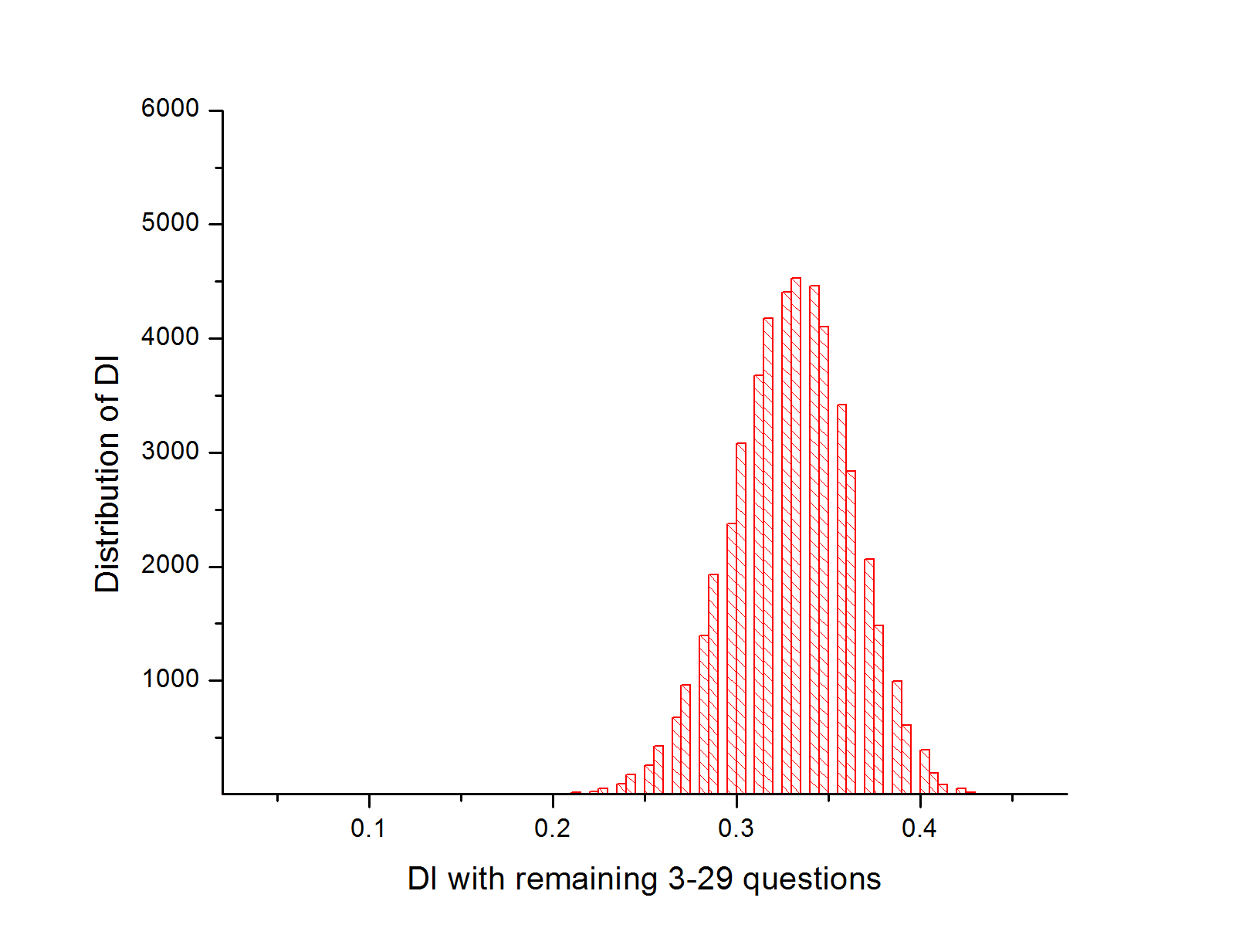}  \\
\hline{}
\end{tabular}
\caption{Distribution of DI values when only combination of remaining 3 to 8 questions are considered for calculating
total. Second plot represents the 3 to 29 combinations}\label{gauss1}
\end{figure}
 there were 497 students who
took the test which had 30 MCQ  and 60 marks worth 12 LAQs. The analysis of these test results were carried out.

For question number 1, it was decided to obtain value of DI for any of
the 3 or 4 combinations of the remaining questions. The distribution
obtained was not Gaussian but a skewed Gaussian as can be seen in
Figure \ref{sk-gauss}.

When the number of combinations are increased from any 3 to 8 of the
remaining questions, the distribution of DI values take up a more
Gaussian shape as seen in Figure \ref{gauss1}. It needs to be
mentioned here that the value of DI obtained are not continuous values
but multiples of $\frac{1}{C ~ M_j}$ ( defined in section \ref{mot})

However, when the combination is increased from any 3 to 29
combinations the distribution is more gaussian in nature seen in
Figure \ref{gauss1}.
The value of average DI (with sigma) for question number 1 are listed in Table \ref{meanDIQ1}.
\begin{table}[h]
\centering
\caption{Table of average DI with standard deviations}\label{meanDIQ1}
\begin{tabular}{|c|c|c|}\hline
Q. range  &  Average  DI                           \\ \hline
3-4 combinations &  0.26 $\pm$ .07  \\
3-8 combinations & 0.31  $\pm$ .04  \\
3-29 combinations &0.33   $\pm$ .03  \\ \hline
\end{tabular}\end{table}

The analysis of all the questions can be seen in Table \ref{tab-allDIs}. The headings of the different calculations, listed in the 
first row of the table are : 
 Diff-I, DI on the basis of all marks (including long-question
marks and negative marking), all MCQ with negative marking, all MCQ
without negative marking, DI without the self marks,( i.e. marks of
the question under consideration are dropped while calculating total),
DI calculated on all possible 3 to 8 remaining questions, DI on all
possible 3 to 29 questions with random selection, maximum possible DI
under any combination and minimum possible DI under any combination. The values in this
table are listed in increasing order of Diff-I.

~~

\begin{sidewaystable}
\caption{Table of all possible DI values for INJSO 2016}\label{tab-allDIs}
\begin{tabular}{|c|c|c|c|c|c|c|c|c|c|c|}\hline
Q.	&Diff-I  &Theory &	Maximum&	DI	&DI	        &	DI               &DI         &	DI with   &	DI with 	& Minimum \\
 No.     &    & limit      & DI                  &    with    &  30 MCQ &  30     & without & 3-8              & 3-29                  & DI \\
           &      &  of DI   &                      & non-MCQ&including   &  MCQ,        & self       & Combinations& Combinations  &  \\
           &               &              &                     & marks     & -ve           &no -ve      & marks   &  Average       & Average          & \\
           &               &             &                      &               & marking    & marking   &              &                      &                         &   \\  \hline
           &               &             &                      &               &                 &                 &              &                      &                         & \\
9	&	0.11	&0.43	&0.15	&-0.01 &0.09	&	0.09	&0.07	&	0.07 $\pm$	0.02	&	0.07$\pm$	0.02	&	-0.01 \\
30	&	0.12	&0.44	&0.14	&-0.04 &0.05	&	0.00	&0.04	&	0.04	$\pm$0.03	&	0.04$\pm$	0.02	&	-0.04 \\
29	&	0.15	&0.57	&0.33	& 0.16 &0.29	&	0.31	&0.24	&	0.21$\pm$	0.04	&	0.23$\pm$	0.03	&	0.07 \\
2	&	0.15	&0.57	&0.19	&-0.02 &0.10	&	0.11	&0.10	&	0.09$\pm$	0.03	&	0.09$\pm$	0.02	&	-0.01 \\
18	&	0.18	&0.68	&0.16	& .05	 &0.14	&	0.07	&0.07	&	0.08$\pm$	0.03	&	0.08$\pm$	0.02	&	-0.02 \\
23	&	0.21	&0.78	&0.25	&0.10	 &0.22	&	0.16	&0.16	&	0.14$\pm$	0.03	&	0.15$\pm$	0.02	&	0.04 \\
27	&	0.21	&0.78	&0.26	&0.06	 &0.21	&	0.14	&0.17	&	0.13$\pm$	0.03	&	0.14$\pm$	0.02	&	0.01 \\
13	&	0.22	&0.82	&0.38	&0.11	 &0.33	&	0.28	&0.28	&	0.24$\pm$	0.04	&	0.27$\pm$	0.03	&	0.11 \\
19	&	0.24	&0.89	&0.33	&0.12	 &0.31	&	0.32	&0.24	&	0.22$\pm$	0.03	&	0.24$\pm$	0.03	&	0.10 \\
24	&	0.25	&0.93	&0.52	&0.24	 &0.51	&	0.48	&0.47	&	0.36$\pm$	0.05	&	0.41$\pm$	0.04	&	0.13 \\
11	&	0.27	&0.99	&0.28	&0.08	 &0.25	&	0.13	&0.16	&	0.18$\pm$	0.03	&	0.18$\pm$	0.03	&	0.07 \\
5	&	0.30	&1	&0.48	&0.18	 &0.45	&	0.40	&0.39	&	0.32$\pm$	0.05	&	0.36$\pm$	0.03	&	0.13 \\
22	&	0.32	&1	&0.48	&0.22	 &0.49	&	0.42	&0.40	&	0.33$\pm$	0.04	&	0.37$\pm$	0.03	&	0.18 \\
7	&	0.33	&1	&0.31	&0.01	 &0.30	&	0.19	&0.23	&	0.20$\pm$	0.03	&	0.20$\pm$	0.03	&	0.10 \\
8	&	0.36	&1	&0.45	&0.13	 &0.45	&	0.39	&0.37	&	0.32$\pm$	0.04	&	0.34$\pm$	0.03	&	0.19 \\
12	&	0.36	&1	&0.62	&0.28	 &0.62	&	0.61	&0.57	&	0.44$\pm$	0.05	&	0.50$\pm$	0.04	&	0.19 \\
16	&	0.38	&1	&0.56	&0.12	 &0.57	&	0.50	&0.48	&	0.39$\pm$	0.05	&	0.43$\pm$	0.04	&	0.18 \\
28	&	0.38	&1	&0.38	&0.07	 &0.38	&	0.25	&0.30	&	0.26$\pm$	0.03	&	0.27$\pm$	0.03	&	0.10 \\
3	&	0.41	&1	&0.39	&0.07	 &0.37	&	0.26	&0.29	&	0.25$\pm$	0.04	&	0.26$\pm$	0.03	&	0.13 \\
14	&	0.45	&1	&0.79	&0.43	 &0.80	&	0.79	&0.72	&	0.55$\pm$	0.08	&	0.64$\pm$	0.06	&	0.18 \\
1	&	0.48	&1	&0.46	&0.16	 &0.44	&	0.25	&0.37	&	0.31$\pm$	0.04	&	0.33$\pm$	0.03	&	0.04 \\
17	&	0.48	&1	&0.63	&0.21	 &0.66	&	0.51	&0.53	&	0.44$\pm$	0.06	&	0.49$\pm$	0.05	&	0.23 \\
26	&	0.49	&1	&0.73	&0.33	 &0.74	&	0.63	&0.61	&	0.51$\pm$	0.07	&	0.58$\pm$	0.05	&	0.15 \\
20	&	0.49	&1	&0.57	&0.27	 &0.57	&	0.45	&0.49	&	0.41$\pm$	0.05	&	0.45$\pm$	0.04	&	0.23 \\
15	&	0.50	&1	&0.76	&0.40	 &0.80	&	0.68	&0.66	&	0.53$\pm$	0.07	&	0.60$\pm$	0.05	&	0.28 \\
21	&	0.51	&1	&0.67 &0.30	 &0.69	&	0.58	&0.59	&	0.46$\pm$	0.06	&	0.52$\pm$	0.05	&	0.23 \\
4	&	0.51	&1	&0.73	&0.34	 &0.78	&	0.63	&0.64	&	0.51$\pm$	0.07	&	0.58$\pm$	0.05	&	0.28 \\
6	&	0.55	&1	&0.56	&0.11	 &0.56	&	0.35	&0.43	&	0.41$\pm$	0.04	&	0.43$\pm$	0.03	&	0.27 \\
10	&	0.67	&1	&0.65	&0.22	 &0.66	&	0.49	&0.56	&	0.49$\pm$	0.04	&	0.54$\pm$	0.03	&	0.30 \\
25	&	0.68	&1	&0.64	&0.13	 &0.68	&	0.38	&0.51	&	0.49$\pm$	0.04	&	0.52$\pm$	0.03	&	0.18 \\ \hline
\end{tabular}
\end{sidewaystable}

\begin{figure}[h]
\includegraphics[scale=0.40]{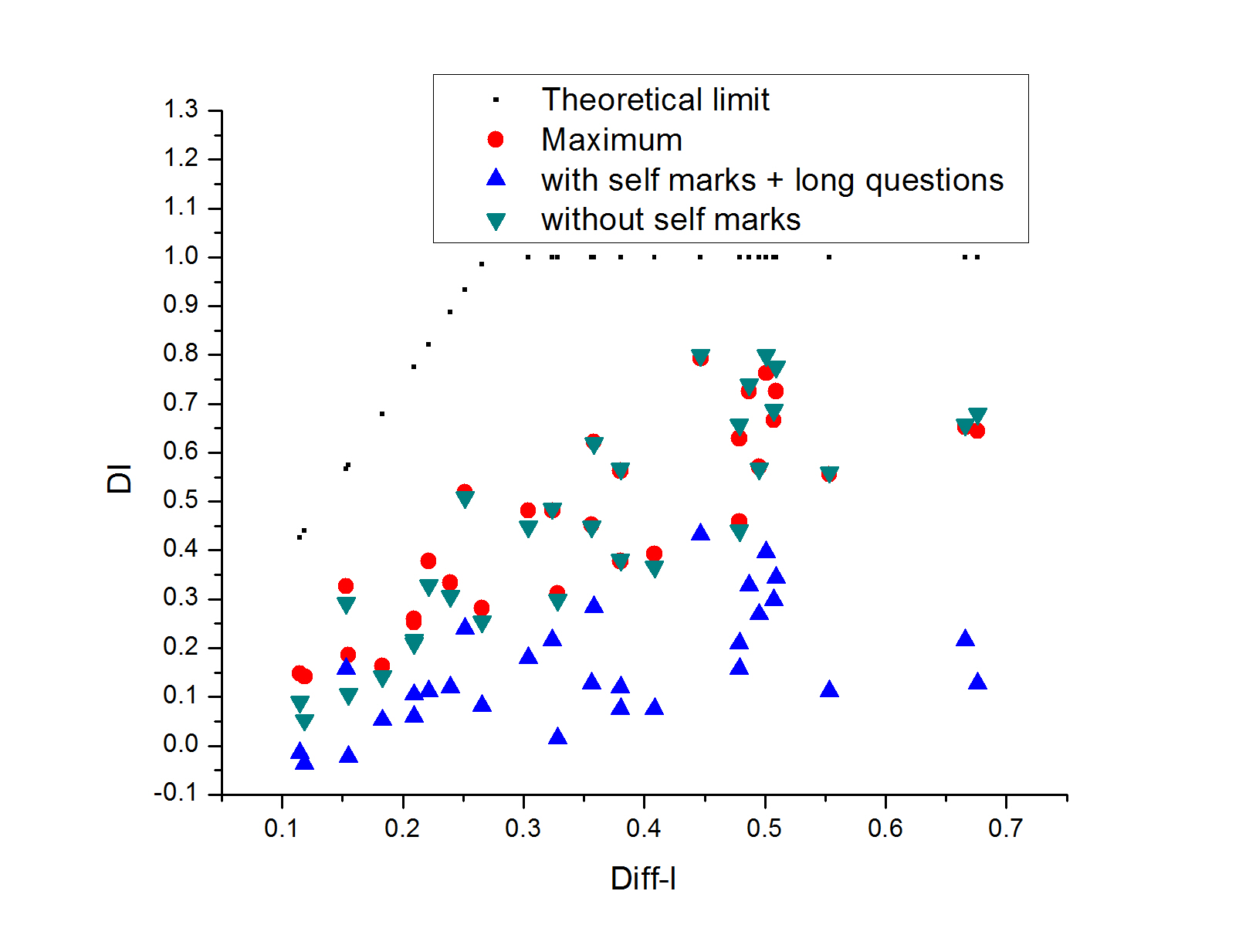}
\caption{Distribution of DI values when only combination of remaining 3 to 29 questions are considered for calculating
total.}\label{di-theo}
\end{figure}

\begin{figure}[h]
\includegraphics[scale=0.40]{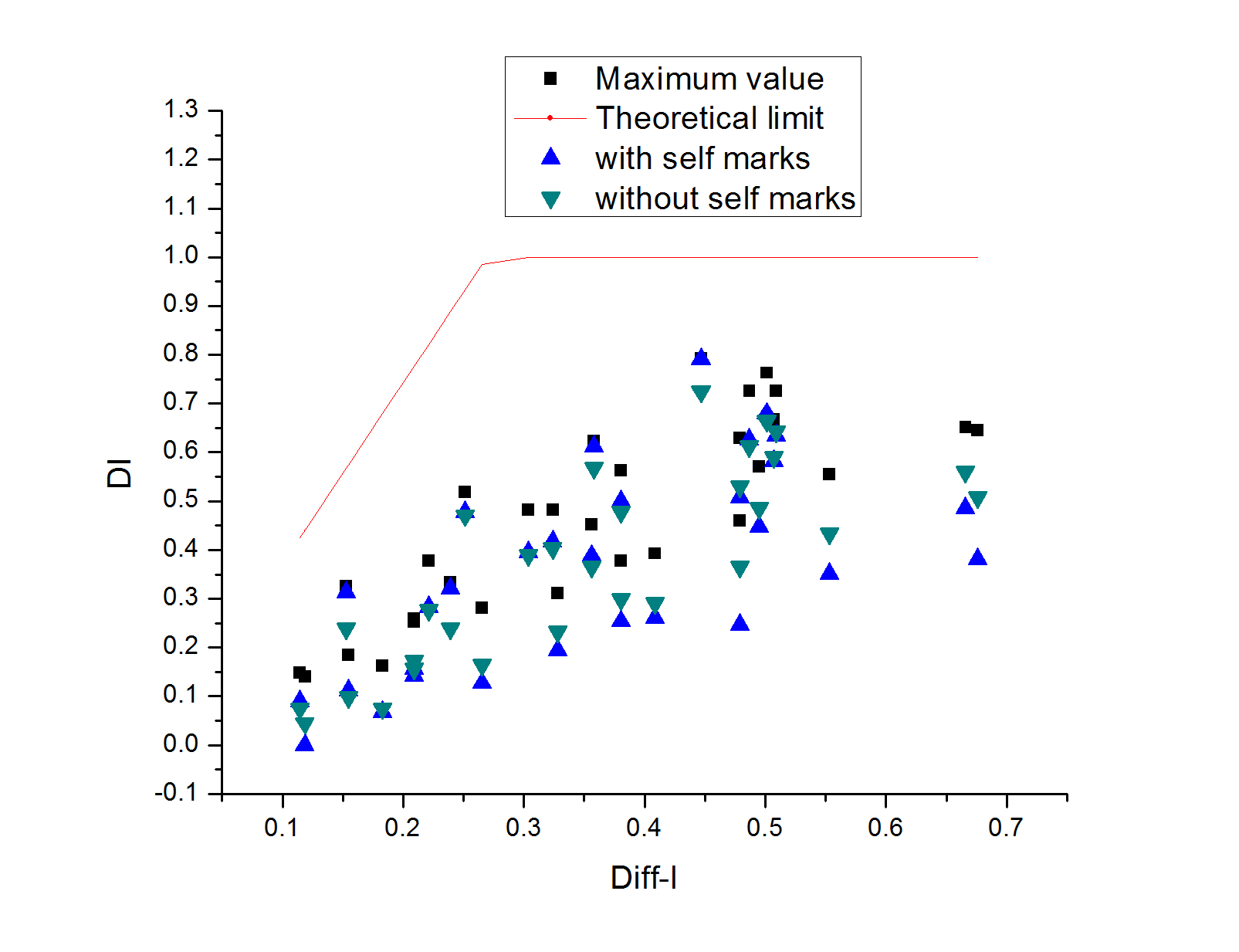}
\caption{Distribution of DI values when only negative marking scheme is ignored and self
marks are also excluded.}\label{di-witho}
\end{figure} 
\begin{figure}[h]
\includegraphics[scale=0.50]{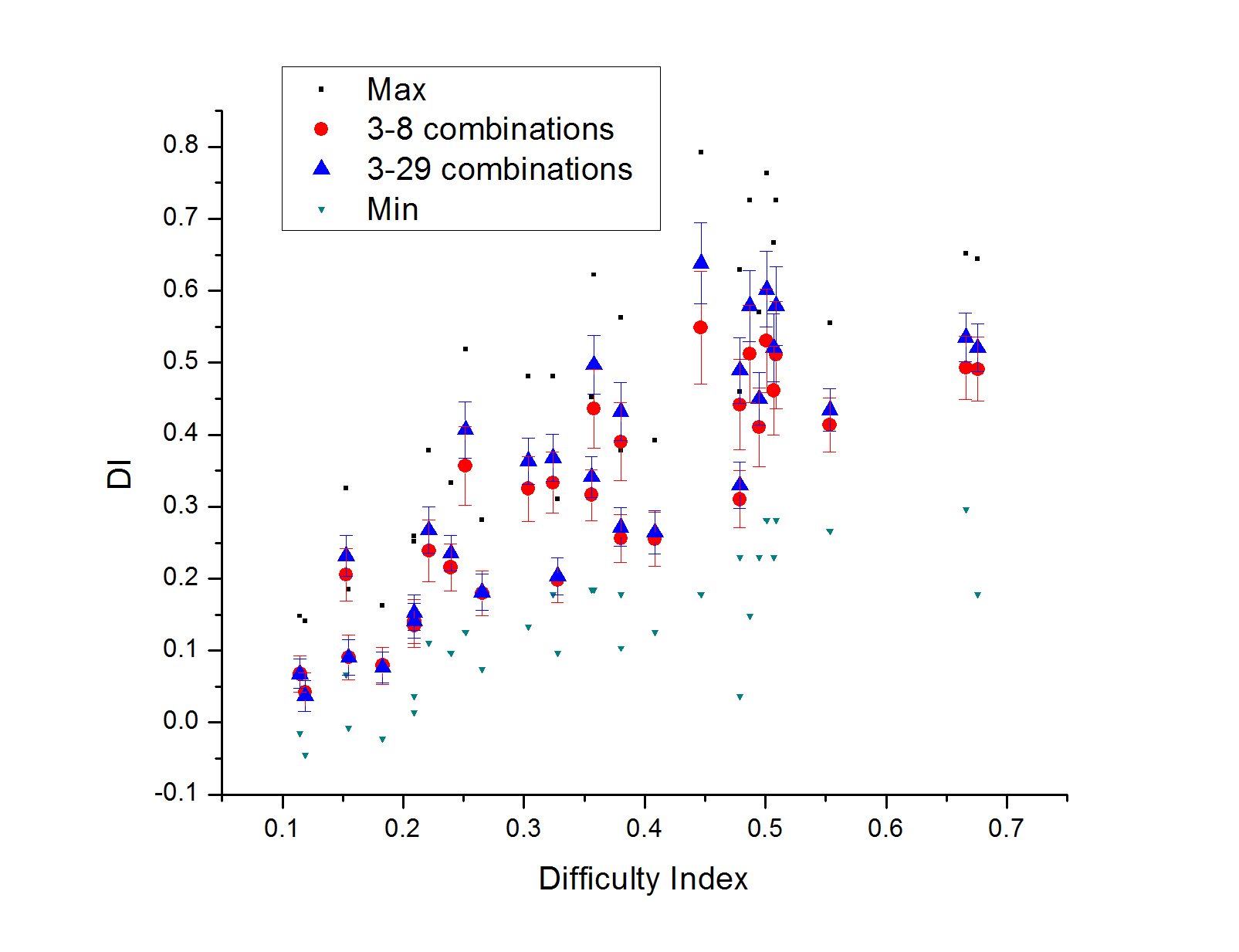}
\caption{Distribution of DI values calculated using 3-8 combinations and 3-29 combinations.}\label{2-ranges}
\end{figure} 
 Figure \ref{di-theo} depict the data are plotted as a function of
 Diff-I.
It can be noted that when all marks are
included the DI values tend to be lower than the DI values obtained
when only MCQ are considered, with their negative marking
included. Those values are very close to the maximum value of DI
obtained in the given data.

Similar trend can also be observed when negative marking is dropped and DI is calculated using all 30 questions
or when the self marks (the marks of the question whose DI is being calculated) are dropped. This can be seen in Figure
\ref{di-witho}.  However, 
it can be seen that the maximum value is always higher than these values. 

Hence, the next logical question is what can be the spread in the values of DI under different possibilities of combinations 
of the questions. A special code was run to calculate all possible combinations of remaining 3  to 8 questions of the 
questions other than the one whose DI is being calculated. 
  A similar attempt was made to determine the spread of DI from all possible 3 to 29 combinations. However, this
runs into a calculation involving more than 10$^8$ trials. Hence a random number operator was introduced to
truncate the basis space of calculations by a factor of 10000. The result from these calculations are depicted 
in Figure \ref{2-ranges}.

The two values agree within the range of standard deviation and hence the 3-29 combination DI values tend to represent the question
more appropriately than when calculated using total of all marks.

In all these figures, thin dots represent the maximum and minimum value of DI possible under various combinations.

\subsection{Data set 3: IJSO-2013}\label{sec-ijso2013}

After having established the reason to report DI values as a range of numbers, represented by an average number along with its error $\sigma$, we extend the idea to use the code for long questions which may be theoretical in nature or arising out of practical exam. The code also handles the negative marks in the multiple choice nature of exam thus making it a handle-all comprehensive code to analyze the questions. Since there were 30 MCQs, 10 LAQs (5 long questions and 5 experimental questions), leading to approx 4 x 10$^{13}$ combinations,  a truncation factor of 10$^7$ was used. 

\begin{figure}
\includegraphics[scale=0.40]{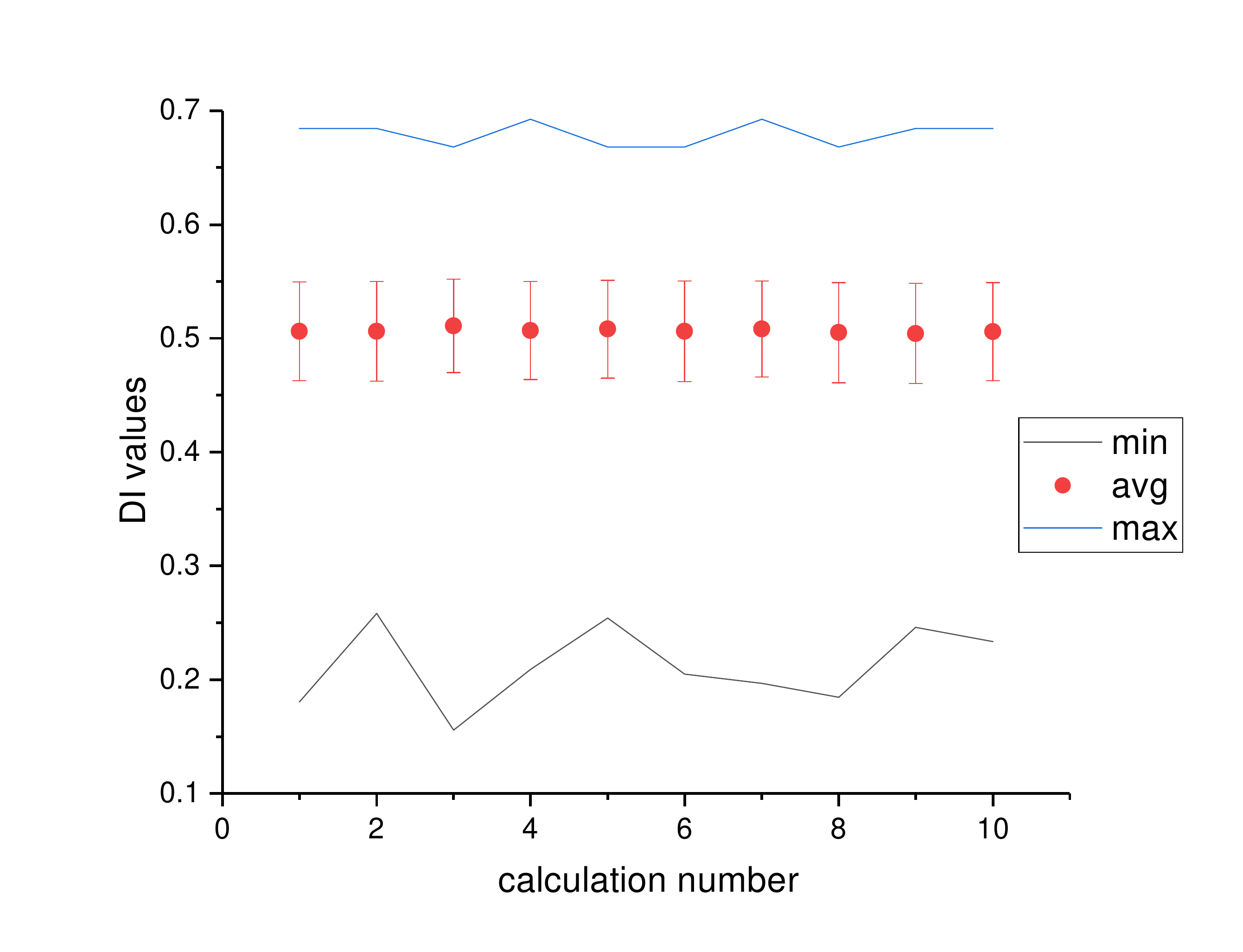}
\caption{Ten trial runs of DI calculations for question number 5.}\label{di-repeats}
\end{figure}

The full set of choices is truncated by a random number, hence in order to see if the DI values calculated on a truncated set represent the full data. The DI values were calculated in ten different trials, for the same question and the results plotted in Figure \ref{di-repeats}.

It can be seen that the variation in the average value of DI($\pm \sigma$) for each of the 10 trials , is much less than the value of $\sigma$.

\begin{figure}
\includegraphics[scale=0.40]{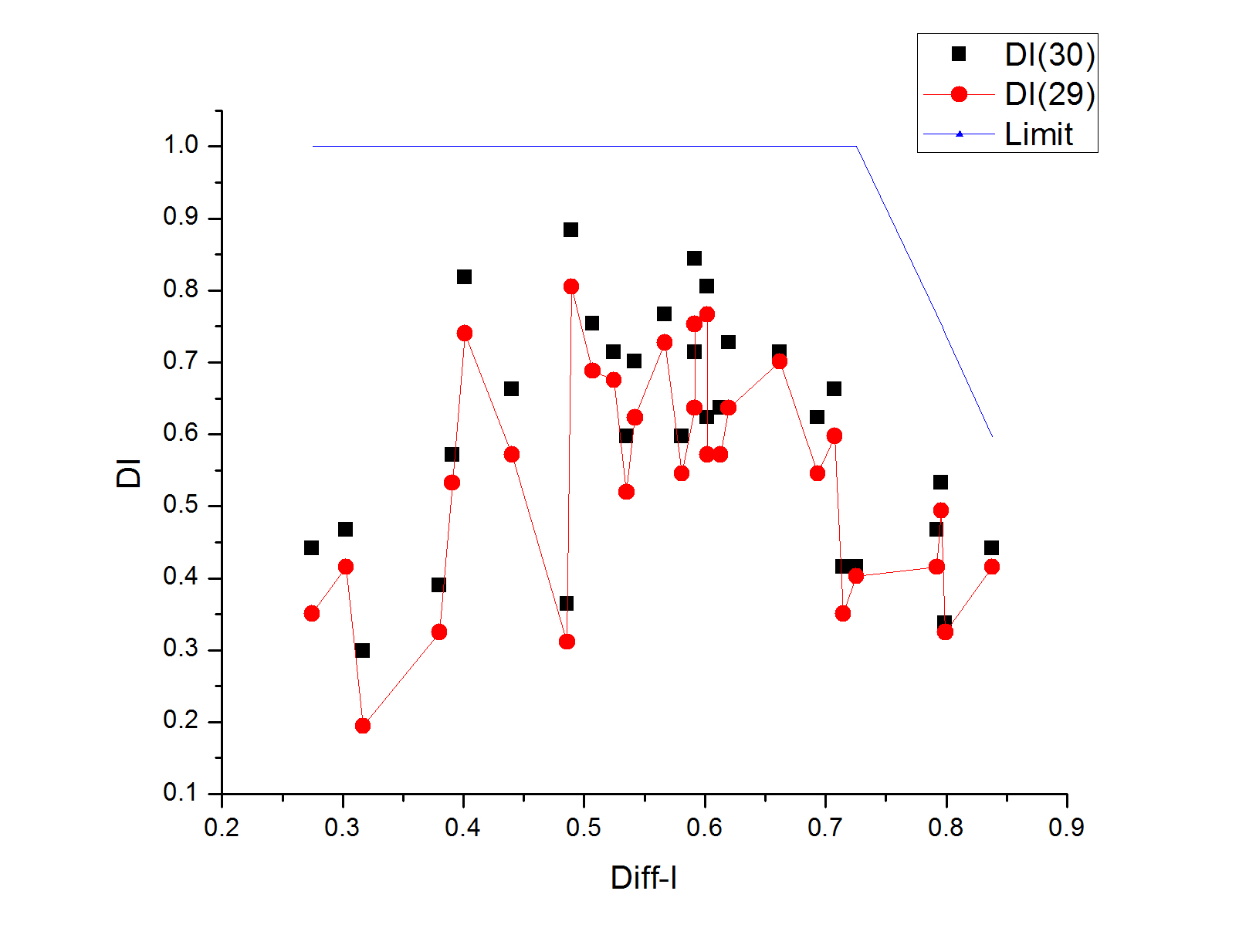}
\caption{Distribution of DI based on 30 (self marks) and 29 questions (without self marks) as 
a function of Difficulty index Diff-I. This is IJSO 2017 international data. Do we need to keep it now?}\label{diffI-di-ijso2017}
\end{figure}

The code analysis conducted on the data from International event held in India in the year 2013 is depicted in the Figure \ref{diffI-di-ijso2017}. The representation clearly shows  the difference in 
DI values when the self marks are used and when not used, to calculate the DI values. For simplicity, only the MCQs are analyzed to display this effect.

\begin{figure}
\includegraphics[scale=0.40]{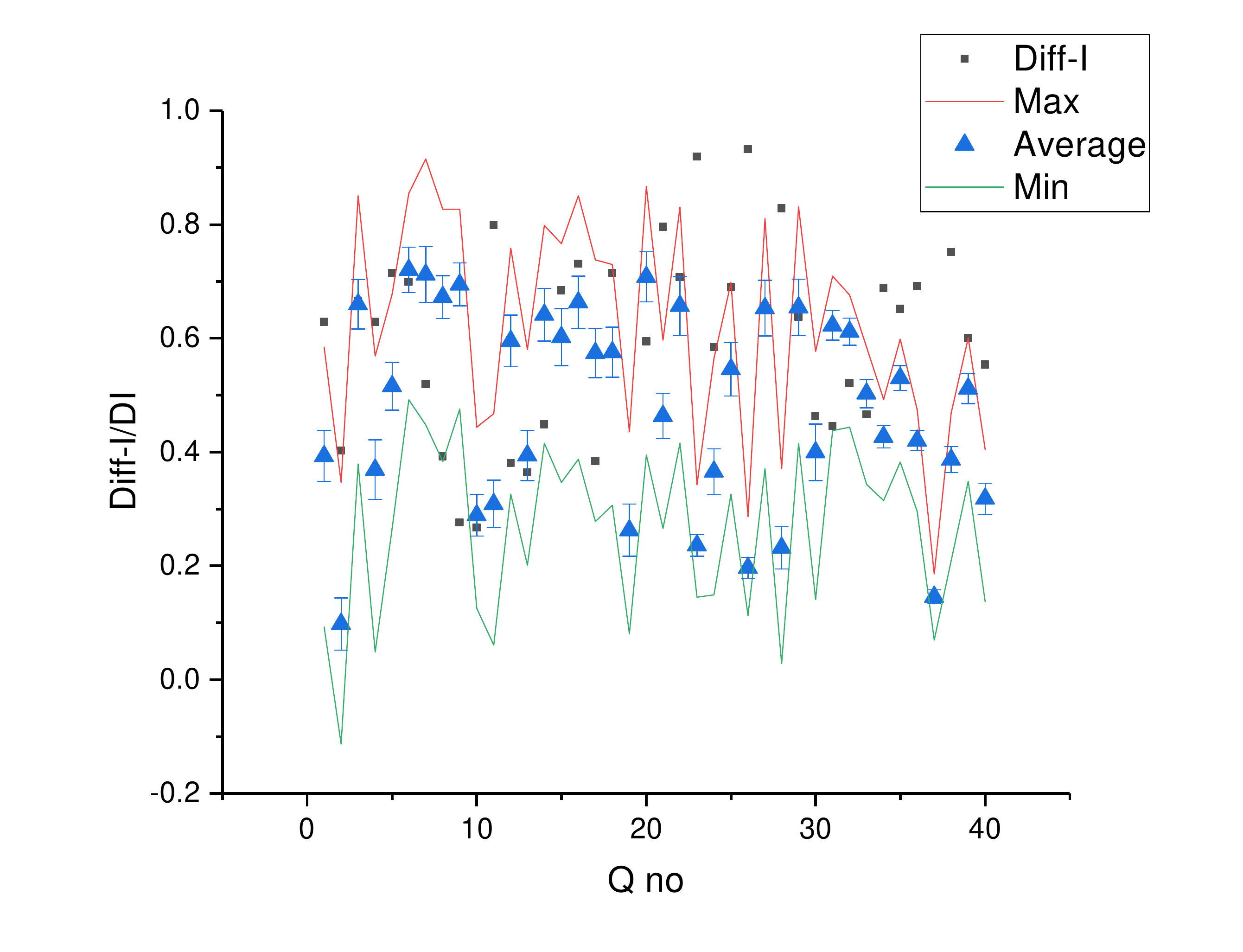}
\caption{Values of difficulty index and DI as a function of question numbers where the last 5 questions represent the experimental questions and the 5 questions before that represent the theory (long problems) questions}\label{di-2013}
\end{figure}

Figure \ref{di-2013} depicts the values of Diff-I and the DI values as a function of question numbers which also include the theoretical and experimental questions.

\begin{figure}
\includegraphics[scale=0.40]{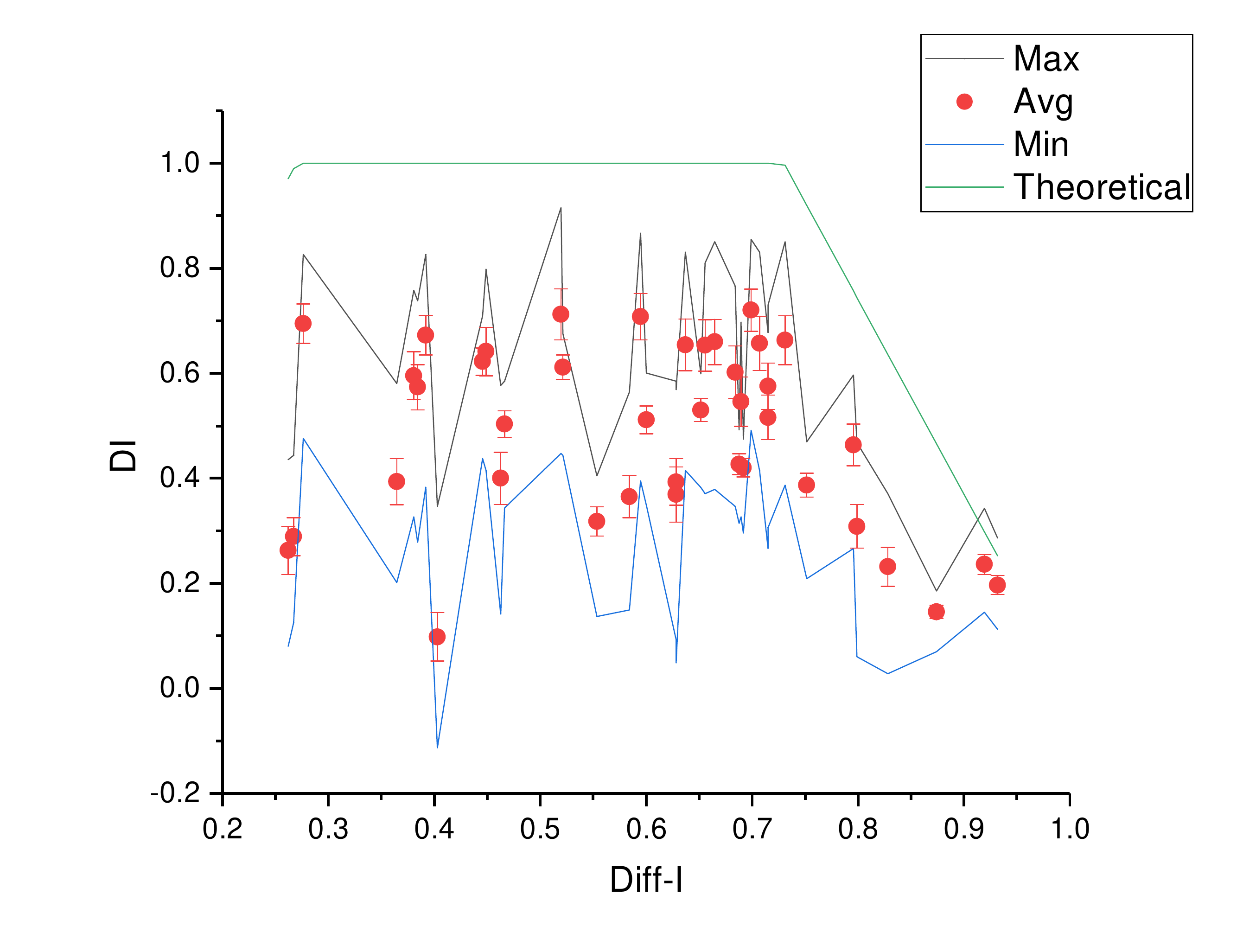}
\caption{Values of difficulty index and DI as a function of question numbers where the last 5 questions
represent the experimental questions and the 5 questions before that represent the theory (long problems)
questions}\label{diff-di-2013}
\end{figure}

In order to study the variation of DI values as a function of Difficulty index the data is plotted differently and can be seen in Figure \ref{diff-di-2013}, for all the questions.

Here it can be seen that the only one question, which happens to be a MCQ, is having a very low DI value and at the same time has Difficulty index of around 0.4, which implies that only forty percent of the students answered the questions  correctly but at the same time did not help discriminate between ``good" and ``not-so-good" students. Similarly an experimental question has very low value of DI but has very high value of Diff-I, implying that almost every student answered this question and hence did not have very high discrimination value. Other than this question, 21 questions had very high values of DI, (ranging between 0.5 to 0.8) and 17 questions happen to have reasonable values of DI (ranging between 0.1 to 0.4). 
  
\section{Conclusions and Implications}
\label{conc}

The analysis clearly shows that the conventional analytical approach does not yield fine grained in outcomes when it comes to 
(a) LAQs, (b) Experimental questions and (c) When no-response trend is influential.
From the values of DI values in Table \ref{tab-di-range} it is clear that, calculating ``total'' using the marks of the 
question in consideration, which we call ``with self'' data, gives values of DI  larger in many cases, than the case where the self marks are not included in calculating the ``total'' for obtaining DI values. If several combinations of remaining questions are used to calculate the value of ``total'', then one gets a range of values of DI which mostly cover the DI value when calculated using different mechanisms. Reporting the DI values in form of a range i.e. average with standard deviation, gives a much broader view of the DI value of a question as compared to a single value from a large possibility of values of DI.

For those questions whose Diff-I values are less than 0.27 or greater than 0.73, the value of DI has a certain ceiling on the values. Those questions which have Diff-I of around 0.5 have the highest potential and possibility to have DI values as close to 1 as possible. Thus, all those questions which have Diff-I close to 0.27 and tend to have maximum value of DI are ideally the best questions. These are the questions which have highest discrimination power with minimum number of participants correctly answering the questions. 

Those questions with high Diff-I (specially those above 0.73), do not contribute much towards filtering students in a competition. But they can be a potentially used for giving a moral boost to students appearing for the exam. 
At the same time those questions which have Diff-I close to 0.5, i.e. having highest probability to have large value of DI, but instead have low DI values, will be the most undesirable questions.

There seem to be a range of question item possibilities that add value, rigour and are helpful in deriving interpretations
from large data set. For instance, low Diff-I but high on DI, low on DI but high on Diff-I need to be carefully examined in 
the larger context of the intended design. The findings from this study open this avenue for further investigation. 

Analyzing question items using the proposed scheme allowed to assess participants performances as well as provided important clue on the quality of questions. The reliability of devised algorithm has been established through its operation on three distinct data sets. 

While data sets used here involve high stake competition as IJSO, the revised algorithm may be tested with other instruments involving high number of participants in various National level tests. While this prospect is proposed, the authors seek
to take forward a  more detailed study of qualitative assessment of each question on the basis of DI values obtained, is the scope
of future work and is in progress.

\section{acknowledgements}
Authors would like to thank Junior Science Olympiad team members for their valuable inputs.  Authors also acknowledge the Government of India for funding the Junior Science Olympiad through Tata Institute of Fundamental Research and the Department of Atomic Energy.

\end{document}